# Progress and Opportunities of Foundation Models in Bioinformatics

Qing Li[1], Zhihang Hu[1], Yixuan Wang[1], Lei Li[1], Yimin Fan[1], Irwin King[1], Le Song[2,*], Yu Li[1,*]

[1]Department of Computer Science and Engineering, Chinese University of Hong Kong, Hong Kong SAR, China
[2]BioMap, Beijing, China

**Abstract**
Bioinformatics has witnessed a paradigm shift with the increasing integration of artificial intelligence (AI), particularly through the adoption of foundation models (FMs). These AI techniques have rapidly advanced, addressing historical challenges in bioinformatics such as the scarcity of annotated data and the presence of data noise. FMs are particularly adept at handling large-scale, unlabeled data, a common scenario in biological contexts due to the time-consuming and costly nature of experimentally determining labeled data. This characteristic has allowed FMs to excel and achieve notable results in various downstream validation tasks, demonstrating their ability to represent diverse biological entities effectively. Undoubtedly, FMs have ushered in a new era in computational biology, especially in the realm of deep learning. The primary goal of this survey is to conduct a systematic investigation and summary of FMs in bioinformatics, tracing their evolution, current research status, and the methodologies employed. Central to our focus is the application of FMs to specific biological problems, aiming to guide the research community in choosing appropriate FMs for their research needs. These downstream specialized tasks include sequence analysis, structure prediction, function annotation, and multimodal integration. In each section, we delve into the specifics of the problem at hand. We compare the structures and advancements of FMs against traditional methods, emphasizing their applications across different biological domains. Furthermore, the review analyses challenges and limitations faced by FMs in biology, such as data noise, model explainability, and potential biases. This analysis provides a theoretical foundation for understanding why certain FMs might underperform in specific tasks. Finally, we outline potential development paths and strategies for FMs in future biological research, setting the stage for continued innovation and application in this rapidly evolving field. This comprehensive review serves not only as an academic resource but also as a roadmap for future explorations and applications of FMs in biology.

Bioinformatics brings efforts to discovering meaningful insights from amino acid sequences, protein structures, single-cell transcriptomics, bio-medical text and images, and other diverse biological data. These efforts facilitate crucial applications such as disease detection, drug design, novel therapy discovery, *etc*., but have limited generalizability and may need substantial customization on specific datasets over decades [Hughes et al., 2011; Bommasani et al., 2021]. On the contrary, artificial intelligence (AI), powered by increasing data availability and computational resources, offers an alternative approach [Topol et al., 2019] to obtain characteristics of biological insights by integrating deep learning mechanisms such as multilayer perceptron (MLP) for nonlinear features [Park *et al*., 2016], convolutional neural network (CNN) for image features [Wang *et al*., 2018], recurrent neural network (RNN) for time series features [Shen *et al*., 2021], transformer for natural language features [Whalen *et al*., 2016], graph neural network (GCN) for features represented as graph [Forster *et al*., 2022], and graph attention network (GAT) targets graph features with distinct attention [Dong *et al*., 2022]. Foundation models focus on pre-training large-scale models from massive data to obtain generalizable features that can be easily adapted to downstream tasks in a fine-tuned, few-shot, or zero-shot manner and can significantly boost performance, thus giving them growing attention and popularity in the field of AI [Mahmud et al., 2018]. Currently, general purpose FMs consist of digital data of different modes pre-trained and finetuned for computer applications of interest (Fig. 1(i)). They have been firmly established as the state-of-the-art approach for question answering [Wiggins et al., 2022], video games [Baker et al., 2022], AI education [Tack et al., 2022], medical AI [Moor et al., 2023], and other applications in computer science.

Recently, foundation models have deciphered immense potential in bioinformatics. A key strength of FMs lies in their capacity to learn dependable representations of intricate biological datasets. This is facilitated by data-intensive pre-training, a process that researchers can easily utilize for various downstream tasks with limited data through fine-tuning mechanisms (*e.g*., transfer learning for varying scale biological targets on a pre-trained foundation model), which makes it easier for researchers to employ pre-trained embeddings acquired from others to solve their targeting problems. In terms of reproducibility, supervised and semi-supervised FMs offer more robust generalizations in the translation of experimental inputs to outputs. Their flexibility and robustness surpass traditional methods, providing advantages for researchers to decipher complex biological systems. Moreover, unsupervised FMs can significantly enhance biological performance by deeply probing a plethora of underutilized resources that are either unannotated or laden with noise [Rao *et al.*, 2021; Forster *et al.*, 2022]. The robust and reliable capabilities, strong exploration and exploitation capacity, and flexible adaptability to diverse downstream tasks make FMs a compelling approach to address biological challenges, *e.g.*, unknown 3D structures or function annotations prediction from sequences [Sapoval *et al.*, 2022], rare disease discovery from limited task-specific data [Theodoris *et al.*, 2023], and other analyses on noise interference data, overlapped data, or other unexplored data in bioinformatics.

*Correspondence to: liyu@cse.cuhk.edu.hk (Y. L.) and songle@biomap.com (L. S.).

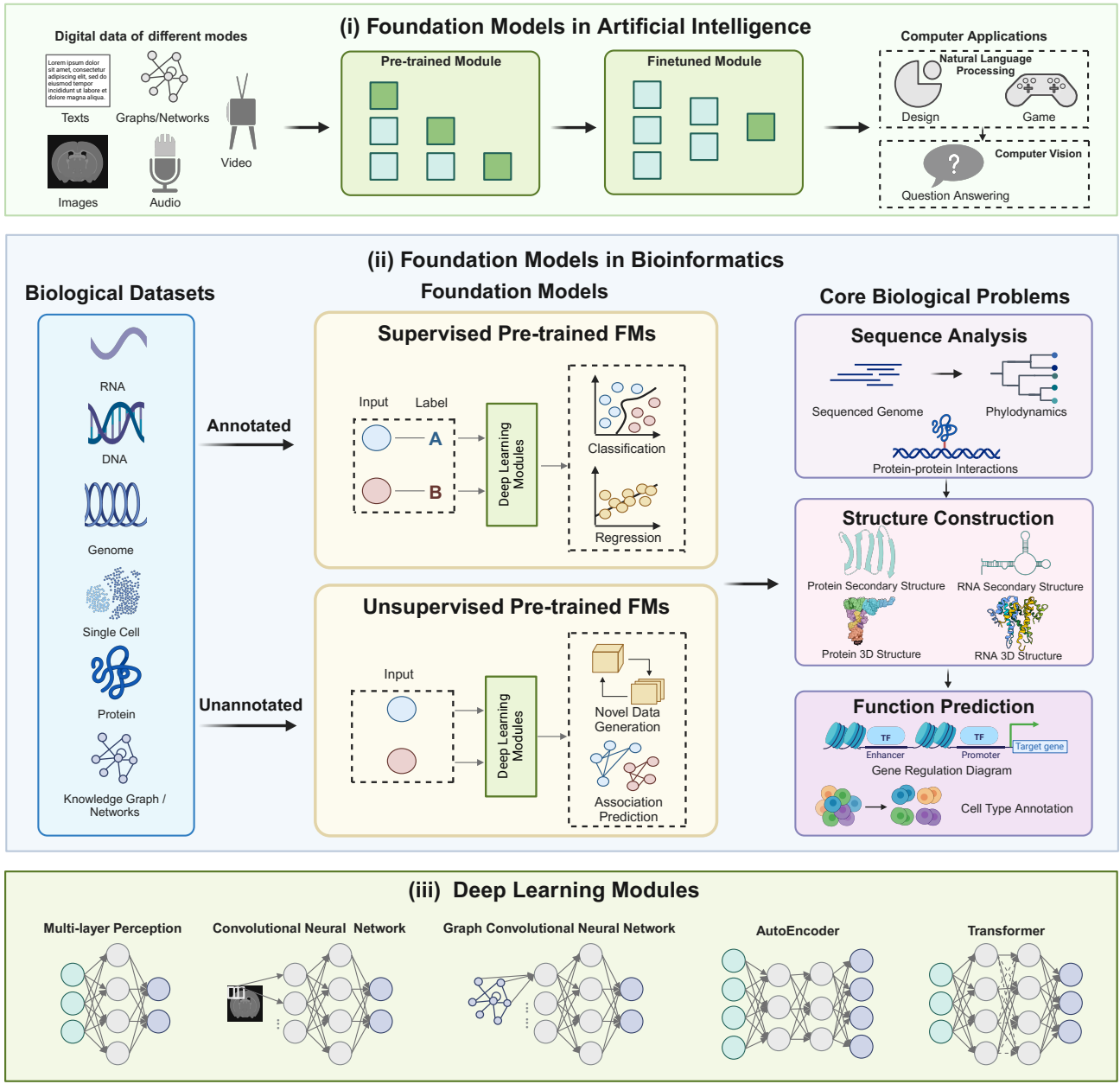

**Figure. 1 Foundation Models (FMs) in Artificial Intelligence and Bioinformatics. (i) Foundation models in artificial intelligence.** General purpose FMs are mainly pre-trained on diverse digital data and finetuned for extensive computer applications such as question-answering systems, image design, and computer games. **(ii) Foundation models in bioinformatics.** FMs in bioinformatics mainly focus on core biological problems including biological sequence analysis, biological structure construction, and biological function prediction on both labeled and unlabeled biological datasets. They can be pre-trained (supervised, semi-supervised, unsupervised) on multiple phases of biological data for various downstream tasks. Based on the pretraining architectures of the foundation model, they can be classified into supervised FMs to capture complex patterns and relationships within the labeled data, and unsupervised FMs to generate new representations not observed in the unannotated training data. **(iii) Deep learning modules.** Deep learning modules are the cornerstone in building deep learning methods, such as multi-layer perception (MLP), convolutional neural network (CNN), AutoEncoder, graph convolutional network (GCN), and transformer (dash arrows represent low attention). Notably, the GCN particularly effective for analyzing undirected graphs has not been employed in FMs. All these deep learning modules can be trained in an end-to-end manner and improve computational efficiency by the parallel processing mechanism.

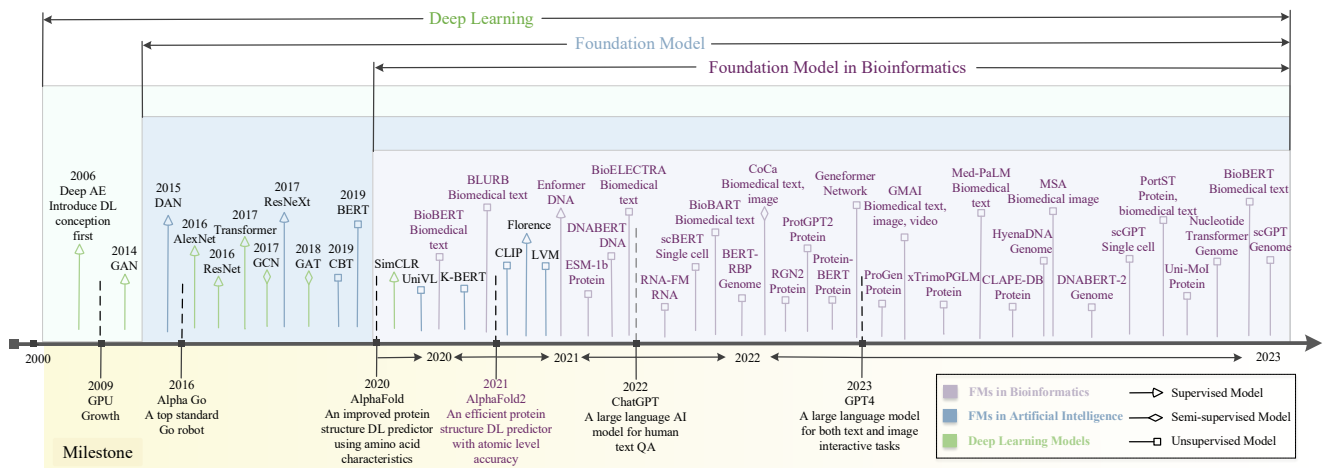

**Figure. 2 Timeline for FMs in Bioinformatics and Their Background in Deep Learning.** FMs in bioinformatics coincided with the rise of deep learning, gaining significant momentum as these models demonstrated remarkable achievements in the era of big data. Pioneering milestones such as Alpha Go, the first robot to meet top standards, significantly enriched the field of deep learning. Subsequent developments, including AlphaFold and AlphaFold2, revolutionized protein structure prediction from biological sequences. The introduction of GPT4 marked a turning point, triggering a surge in the application of FMs. These efforts boosted FMs (supervised FMs, semi-supervised FMs, unsupervised FMs) in bioinformatics showing promising results in solving domain exploration, sequence construction, function prediction, and multimodal integration to acquire salient biological sequence, structure, function, and other complex information for practical applications in biology.

**Table 1 Biological Problems and Their Associated Data in Biological FMs.** The table provides an overview of five distinct problems in bioinformatics solved by biological FMs: domain exploration, sequence analysis, structure construction, function prediction, and multimodal problems. These problems involve one or more categories of biological data, including DNA, RNA, proteins, single cell genomics (scGenomics), knowledge graphs/networks, and biological text/images. Domain-specific problems primarily focus on biological text/images/video. Core biological problems (sequence analysis, structure construction, function prediction) involve genes and mutations, biological phenomena data, and their relation and interactions. Multimodal integration biological problems may use multiple data types, e.g. biomedical text/images and proteins.

| Problems / Data | Domain exploration | Sequence analysis | Structure construction | Function prediction | Multimodal problems |
|---|---|---|---|---|---|
| DNA | | ✓ | | ✓ | |
| RNA | | | ✓ | ✓ | |
| Protein | | ✓ | ✓ | ✓ | ✓ |
| scGenomics | | ✓ | ✓ | ✓ | ✓ |
| KGs/Net. | | ✓ | | ✓ | |
| Text/Image | ✓ | | | ✓ | ✓ |

Although FMs have been widely employed in many research fields and industry applications, there is still a lack of a thorough comprehension of FMs in bioinformatics. Indeed, previous approaches commonly rely on general domains of text/image/video/graph to analyze targets with the help of natural language processing, image processing, or graph learning methods [Zou *et al.*, 2019]. Unfortunately, the utility of these methods is often constrained for biological researchers. Typically, there exist at least three critical challenges associated with the application of FMs in AI for addressing biological problems.

Firstly, general purpose FMs that are mainly constructed by task-specific approaches for a particular downstream task without expertise requirements often lead to notoriously model overfitting in bioinformatics. Secondly, most FMs rely heavily on large-scale training datasets, making them inflexible and untrusted when analyzing shifted datasets for significant biological applications [Uhlmann *et al*., 2022]. As depicted in the futility theorem, the predicted results of binding sites may not be functional in vivo, even though they bind to the vitro sequence with higher possibility [Wasserman *et al*., 2004]. At times, specific biological problem-solving necessitates the incorporation of specialized insights derived from knowledge in the biological domain. Finally, non-linear deep features derived from FMs in AI for biological tasks may encounter challenges in terms of biological interpretability and model reliability, owing to their intricate structures and the diverse nature of biological targets.

In this context, a review that encapsulates the cutting-edge applications of FMs in bioinformatics is valuable and necessary to bridge the existing gap in a comprehensive understanding of these models, shedding light on their operational mechanisms, the contributions they have made, and the challenges and opportunities they present for future explorations. Macromolecules are fundamental to addressing core biological problems, including sequence analysis, structure construction, and function prediction. In response to these problems, we provide a comprehensive survey of foundation models in bioinformatics. These FMs are trained in a supervised or unsupervised manner and can be applied to downstream applications like domain exploration, core biological problems, and multimodal integration biological problems. These problems are intricately linked with biological data: DNA, RNA, proteins, and single cell genomics, as well as knowledge graphs/networks, and text/image data, shown in Table 1. Additionally, the evolution of FMs in bioinformatics presented in Fig. 2 emphasizes their development of network structures in deep learning, general purpose FMs, and FMs in bioinformatics with significant milestones.

This review delivers a thorough understanding of recent developments and challenges associated with foundation models that emphasize a sound grasp of biological issues. Except for the domain shift foundation models from general domains to biological domains, they primarily focus on three core biological problems: sequence analysis, structure construction and prediction, and function prediction, which play a crucial role in analyzing the sequence, structure, and function of biological targets. Multimodal integration biological problems, such as multi-modality analysis, may involve multiple types of biological data to further enhance their performance. The review concludes by discussing potential directions in light of current challenges and opportunities. Overall, we review recent foundation models in bioinformatics through the following subsections (i) foundation model architectures, (ii) biological foundation models for five kinds of biological problems on top of introducing distinct problems and datasets, data preprocessing, and down-stream tasks, (iii) challenges and opportunities, (iv) conclusions.

## Foundation model architectures

Foundation models (FMs) are large, pre-trained AI systems that can be effectively generalized across a wide range of domains and tasks [Bommasani *et al.,* 2022]. The advent of general purpose foundation models has been most prominently observed in the field of natural language processing (NLP) [Dai *et al.,* 2015, Howard *et al.,* 2018, Dong *et al.,* 2019]. This development has subsequently permeated into computer vision and other areas of deep learning [Long *et al.,* 2015, Xie et al., 2017, Yuan et al., 2021]. In bioinformatics, FMs trained on massive biological data through supervised, unsupervised, or semi-supervised machine learning strategies offer an unprecedented prediction ability via finetuning mechanisms with limited data. Based on pretraining strategies, FMs in bioinformatics can be divided into two main categories: supervised FMs that construct specific embeddings to capture complex patterns and relationships within the labeled data, and unsupervised FMs that mainly concentrate on various downstream tasks to generate new representations not been observed in the unannotated training data. Notably, some FMs composed of both supervised and unsupervised pretraining mechanisms noted as semi-supervised FMs are elucidated in the supervised pre-trained FMs. Moreover, FMs could also be classified into discriminative and generative models according to the model task. The architectures of the supervised pre-trained FMs and unsupervised pre-trained FMs are introduced as follows.

**Supervised pre-trained FMs.** Traditional AI-based models focus on training different types of neural networks to extract and identify insightful features of training data. They usually train the model in an end-to-end process to solve only one kind of task each time, such as classifying COVID-19 from chest CT images in a supervised learning process [Pathak *et al.*, 2022]. In this context, supervised learning that necessitates preliminary human intervention for the appropriate labeling of input and output data in linear classifiers, support vector machines, decision trees, random forests, etc., is essential for understanding complex biological processes in normal and disease states, which in turn facilitate therapeutic target identification and biological problem-solving. Based on these machine learning mechanisms, supervised pre-trained FMs pre-train the model accurately for supervised classification or regression tasks. For instance, Enformer predicts promoter-enhancer interactions directly from DNA sequences with a large amount of information flow in CNN [Avsec *et al.*, 2021].

Semi-supervised learning techniques, such as autoencoders, contrastive learning, and self-training, are conventionally trained on data sets with a limited number of annotations. For example, as an image-text semi-supervised foundation model, CoCa pre-trains both visions with zero-shot image classification and vision-language with contrastive objectives on noisy image-text pairs by dual-encoder pretraining models [Yu *et al*., 2022]. The divergence between supervised and semi-supervised pre-trained models is primarily determined by the volume of annotations necessary to attain the anticipated results. Nevertheless, both supervised and semi-supervised pre-trained FMs possess the capability to update and pre-train neural networks via a back-propagation pipeline, leveraging the statistical outcomes of target variables and their estimated counterparts.

**Unsupervised pre-trained FMs.** Unsupervised learning can directly identify unlabeled input data with human intervention only for the output validation. For example, MPNet provides a permuted language modeling that inherits the merit of masked language modeling and maintains the dependency among predicted tokens [Song *et al*., 2020]. BioKG builds a biological knowledge graph compiled from open biological databases to extract a unique set of relations and support relational learning model tasks [Walsh et al., 2020]. Thereby, unsupervised pre-trained FMs enable dimensionality reduction for more information representation, or more accurate association generation to support more restrictive downstream tasks [Moor *et al*., 2023].

The primary tasks of unsupervised FMs can be directed into two categories: generation/recovery tasks, and understanding tasks. Generation tasks targeting generate novel data with certain properties from unannotated inputs, e.g., ProtGPT2 [Ferruz *et al.,* 2022] generates protein sequences that exhibit amino acid and disorder properties comparable to those found in natural proteins, yet remain distinct from the existing protein space. Recover tasks can filter the noise of corrupted data and recover their original ones, e.g., ProteinBERT [Brandes *et al*., 2022] recovering the uncorrupted data from the received corrupted inputs performed by randomly replacing tokens and adding random false annotations to force the model to predict annotations from sequence alone. Understanding tasks aims at predicting correlations and associations of inputs such as contact prediction. For instance, DNABERT [Ji *et al*., 2021] is pre-trained on non-overlap splitting and random sampling human genome data where 15 percent of k-mers are masked in sequence during the first 100k steps and 20 percent in the rest 20k steps.

The conventional progress of unsupervised FMs is implemented via masking strategies in pretraining. The xTrimoPGLM [Chen *et al*., 2023] employs four distinctive strategies of masking to redesign the sequence of complementarity determining region 3 (CDR3): CDR3 short masking, whole masking, random mutation, and random retrieval. ProtST [Xu *et al*., 2023] fusions both mask prediction and representation alignment into the pretraining task to model the paired protein sequences and text descriptions. Unsupervised foundation models can adaptively manage large amounts of heterogeneous targets. While they behave like generative models during the training phase, they are commonly used as discriminant models in finetuning for downstream tasks whose goal is to predict the label for a given input.

Compared with the application of the supervised learning method AlphaFold on Protein Data Bank (PDB) data achieved a high degree of accuracy [Senior *et al*., 2020], AlphaFold 2, an unsupervised foundation model, has taken this a step further by incorporating a technique similar to noisy student self-distillation with an enhanced level of accuracy. This milestone comes with emerging unsupervised pre-trained FMs highlighting their employment in bioinformatics. For instance, CLAPE-DB combines a pre-trained protein language model and constructive learning to predict DNA binding residues in an unsupervised manner [Liu *et al*., 2023]. With the merit of exploring long-range unknown classes, unsupervised pre-trained FMs can generate meaningful associations of genes that partake in a variety of functional relationships, like shared bioprocess annotations and co-complex participations. HyenaDNA uses a sequence length scheduling technique to stabilize model pretraining and leverages longer context to adapt to novel tasks [Nguyen *et al*., 2023]. Remarkably, the choice of pretraining strategies holds paramount importance for obtaining optimal performance overshadowing the need for innovative model architecture [Azher *et al*., 2023].

## Foundation models for biological problems

To implement FMs in bioinformatics appropriately, biological problems and datasets together with relevant data preprocessing and downstream tasks in biology will be elucidated at first. As this review concentrates on biological macromolecules (including DNA, RNA, protein), single cell genomics, knowledge graphs/networks, and text/images, foundation models illustrated in this part are generally employed to solve problems in macromolecule biology. Along with pretraining architectures, foundation models are classified into supervised models, which capture intricate patterns within annotated data, and unsupervised models, which generate novel representations unseen in the unannotated training data. In the realm of bioinformatics, we introduce foundation models as versatile tools capable of addressing practical biological problems, including domain exploration, sequence analysis, structure construction, function prediction, and multimodal integration. Each of these areas represents a unique challenge within the field of bioinformatics, and the application of FMs provides innovative approaches to these complex issues. Recent FMs in bioinformatics are summarized in Table 2.

**Biological problems and datasets.** FMs can solve practical biological problems that fall within five categories: domain exploration, sequence analysis, structure construction, function prediction, and multimodal integration. Core biological problems are sequence analysis, structure construction, and function prediction. Sequence analysis obtains salient gene information, such as the information of binding sites (commonly encoded as position weight matrices (PWMs)), protein-protein interaction (PPIs), gene expression, etc., from gene and mutation sequence data DNA, RNA (in which each of four kinds of nucleotides is encoded as a one-hot vector like [1, 0, 0, 0]), protein, and genome (the complex of all the genetic information of an organism). Indeed, the information obtained from these models can be utilized to analyze various downstream tasks. For instance, gene expression embodies the functional regulation process of cells. The differences observed between single cell genomics pave the way for the discovery of new cell types. Similarly, Protein-Protein Interactions (PPIs) and their analogs (such as protein-nucleic acid interactions, protein-ligand interactions, protein-small molecule interactions, etc.) encapsulate the physical binding information between them, providing valuable insights into their interactions. Sequential data are often included as annotations of higher-level data that further contain their synergistic or catalytic interactions.

Structure construction focuses on predicting the structures of proteins and RNA from the secondary structure to the quaternary structure based on the primary structure linear sequences of amino acids in a peptide or protein; the secondary structure contains $\alpha$-helix, $\beta$-sheet with three strands, $\beta$-bend, $\Omega$-loop, random coil architecture, and topology targets; the tertiary structure has hydrogen bonds, hydrophobic interactions, and tertiary contacts; and the quaternary structure represents complex molecule structure. As these structures can be represented by statistical information among amino acid residues [Senior *et al*., 2020], many structure construction efforts are made on amino acids in DNA, single cell genomics, and homologous protein families. Moreover, biological sequence data with different positions may have different functions. In this context, they can also be categorized under multiple sequence alignment (MSA) [Sapoval *et al*., 2022].

Function prediction related to biomedicine enables understanding functions of targets such as proteins and variants to predict polypharmacy side-effects, etc. Core biological data for solving this problem are proteins, individual genes, and their spatial interactions commonly encapsulated within knowledge graphs or networks. These networks represent various information indicated as gene interaction networks, disease pathways, patient networks, and networks that capture similarities between cells. Notably, the prediction of biological function is intrinsically linked to the outcomes of gene expression analysis, given that protein functionality is influenced by the degree of gene expression.

Domain exploration involves parsing biological problems by transforming the principles of natural language analysis and computer vision domains into biological areas. Hence, biomedical text like BioBERT [Lee *et al*., 2020] and 2D and 3D biomedical images such as microscopy images [Uhlmann *et al*., 2022] make up major data in solving domain-specific problems. Multimodal integration biological problems can map multiple types of biological data encompassing multi-omics and morphological histology data, etc.

**Data preprocessing.** Data preprocessing is paramount to ensure satisfactory performance before building a model. Original biological datasets may contain multiple inconsistencies caused by varying purposes and acquisition technologies preventing them from being analyzed directly. Adoption of appropriately curated and preprocessed data without exorbitant data overlap, data deficiency, noise interference, or other unexplored data can improve model computational efficiency and representative ability with better model performance. Examples include doublet removal (without duplicate articles), cell-cycle variance removal, data imputation and denoising, dimensionality reduction (reducing the sequence similarity), representation learning, batch effect removal, normalization, background correction, etc.

Doublet removal can avoid mapping duplicate overlapping data with different identifiers or different data that share the same identifying barcode, which plays a significant role in constructing a unique set of relations between entities [Walsh *et al*., 2020, Bernstein *et al*., 2020]. Cell-cycle variance removal focuses on removing vain variations in gene expression between cells emerging along the cell cycle by subtracting out the cell cycle influence [Brendel *et al*., 2022]. It is intractable for data imputation and denoising because only 6-30% of values can be captured under different chemistry versions and depths, and to decipher "true" and "false" (called "dropout") zeros in more than 70% missing values is a guarantee for further identification. To a certain degree, these data can be refined by leveraging similarities with other datasets, or through the construction of multiple sub-neural networks for imputation [Arisdakessian *et al*., 2019]. Dimensionality reduction of wide gene expression profiles represented in high feature dimensions, also known as representation learning, aims at the construction of embeddings that facilitate the identification of data elements. Systematic variations specific to each batch tend to raise challenges in data integration and lead to significant data wastage. Tran *et al*. compared benchmarks of batch effect removal methods such as seGen (variational auto-encoders neural network model and latent space), Scanorama (mutual nearest neighbor and panoramic stitching), and MND-ResNet (residual neural network for calibration) to effectively reduce the variations and batch effects in data captured with different times, types of equipment, or technologies [Tran *et al*., 2020]. Therefore, customized fine-tuning can correct sequencing batches from multiple datasets [Cui *et al*., 2023]. Protein data can be compared on a common scale by normalization to adjust the measurements [Clement *et al*., 2021], and background correction aims to correct for any background noise in the protein data [Mowoe *et al*., 2022].

With preprocessed biological data, the data analysis model can be efficiently employed and mitigate or even eliminate obstacles in biological tasks such as doublet detection and cell-cycle variance annotation. As a result, the judicious utilization of biological data and their corresponding embeddings can significantly enhance the performance of downstream tasks.

**Downstream tasks.** In bioinformatics, the analysis of downstream tasks is permitted to evolve through the application of fine-tuning strategies that are desired for accurate performance in analyzing biological problems of interest based on pre-trained biological knowledge in FMs. Fine-tuning can greatly reduce computational time and barriers to their implementation and is capable of solving biological tasks related to sequence analysis, structure construction, function prediction, domain exploration, and multimodal integration.

For sequence analysis, besides traditional sequence alignment analysis [Hong *et al.*, 2021], homology detection [Steinegger *et al.*, 2019], and molecular evolutionary genetics analysis (MEGA) tasks [Stecher *et al.*, 2020], there are promoter interaction prediction, enhancer-promoter interactions prediction, variants identification, variant effect prediction, signal peptide prediction, gene dosage sensitivity predictions, genetic perturbation prediction, protein understanding, DNA replication, stability prediction, etc. Promoter prediction identifies promoter regions of motifs in transcription start sites of genome-wide sequences. Non-promoter region samples can then be constructed by shuffling and keeping different parts of split promoter sequences with matching lengths. Enhancer-promoter interactions (EPIs) prediction is essential in cell differentiation and can interpret non-coding mutation with potential pathogenicity [Chen *et al.*, 2022]. EPIs are determined by chromatin conformation and, thereby can be inferred by chromatin conformation capture-based (3C-based) techniques or other genetic approaches. In addition, the promoters and enhancers are also known as initial and distal regulatory elements respectively [Novakovsky *et al.*, 2023].

Variant identification discloses human diseases and traits by distinguishing casual from non-casual variants [Avsec *et al.*, 2021]. Variant effect prediction focuses on determining functional important variants and giving priorities to them [Dalla-Torre *et al.*, 2023]. Signal peptide prediction is a binary protein sequence analysis that predicts their presence and locates their cleavage sites [Brandes *et al.*, 2022]. Gene dosage sensitivity predictions present genes that are sensitive to changes in their dosage interpreting copy number variants in genetic diagnosis [Theodoris *et al.*, 2022]. Genetic perturbation prediction aims to forecast perturbed original values or perturbed gene expression values in certain tasks [Cui *et al.*, 2023]. Protein understanding requires accurate representation at the residue level or protein level to understand biological information encoded within proteins [Chen *et al.*, 2023]. The process of DNA replication is governed by specific initiation and termination sites, with the function of the origin of replication being modulated by epigenetic factors. This intricate process can be studied at a population level by leveraging non-transformed, highly proliferative, and karyotypically stable pluripotent stem cells [Ding *et al.*, 2021]. Stability prediction calls for statistical representations of protein informatics such as natural language-inspired representations [Madani *et al.*, 2023].

Structure construction commonly performs secondary or tertiary structure prediction in downstream tasks. Secondary structure prediction was originally achieved by thermodynamic and alignment methods to determine the homologous sequences and their alignments [Chen *et al.*, 2022]. 3D structures, by contrast, need further exploration due to the lack of 3D structure data, which may be constructed on the raised deep learning method. Moreover, other tasks related to DNA, RNA, protein, and genomics such as predicting DNA binding residues, protein-RNA binding preference, protein-ligand binding pose, splicing junction prediction, neuropeptide cleavage, genome structure and evolution, gene network, etc., underlie the discovery of their structure information as well. Predicting DNA and RNA binding proteins is essential for analyzing genetic variants [Alipanahi *et al.*, 2015]. Transcription factors (TFs) are binding proteins in regulate gene expression that can bind motifs (specific DNA sequences) to regulate transcription. Generally, pathogenic functional variants in complex neurodegenerative diseases occur with the change of TF binding intensities [Wang *et al.*, 2018]. Protein-protein interaction prediction aims at revealing bindings between proteins with transient or stable physical connections. Protein-small molecules and protein-nucleic acid interactions are significant prediction tasks that dominate organism activities [Liu *et al.*, 2023].

Splicing junction prediction is crucial for protein synthesis and genetic disease identification, whose variant effects can be predicted with the integration of process-specific scores [Rentzsch *et al.*, 2021]. Neuropeptide cleavage is one of the post-translational modification binary prediction tasks where the maturation of neuropeptides occurs associated with molecule variability for behavioral and physiological states [Brandes *et al.*, 2022]. Genome structure represents genome regulatory element secondary structures, and evolution denotes the evolutionary trend of virus variants [Chen *et al.*, 2022]. Gene network prediction can map networks based on learned connections between genes. Recently, a transfer learning method has been proposed to learn the connection with limited task-specific data showing a promising analysis for rare diseases [Theodoris *et al.*, 2023].

Function prediction captures various properties of RNA/protein/gene functions, discoveries (novel) cell type, functional genetic variants, and functional modules, and describes gene expression regulation, in silico treatment analysis, fitness landscape inference, trajectory inference, etc. Functional properties prediction performs the classification of RNA/protein/gene into several functional groups. For instance, Gene function prediction, from classifying gene and protein functions to analyzing genome-wide experimental data with multiple statistical tests, relies on the coverage and accuracy of the annotation data such as Gene Ontology (GO) annotation data [Mi *et al.*, 2019]. Cell type annotation describes heterogeneity in tissues following cell clustering for further investigation insights into biology and pathology [Cui *et al.*, 2023]. Functional genetic variants identification probes functional variants located inside regions of interest and subsequently repeated prediction with altered alleles [Ji *et al*., 2021]. Functional module detection inputs from networks and functional features to protein complexes and evaluates the overlap of the predicted module and known complex [Forster *et al*., 2022].

Gene expression regulation models a biological process where the genetic blueprint within a gene is harnessed to synthesize a functional product. Chromatin state analysis is commonly used for detecting annotation and regulation of the genome and for further nucleosome-level function prediction with gene expression and other related data [Ernst *et al*., 2011]. Gene expression profile facilitates therapeutic discovery through gene expression similarities measured by distance metric and clinical importance evaluating a certain gene on the gene expression level, e.g., finding a tumor gene compared with normal groups [Tang *et al*., 2017]. In silico treatment is applied to model human disease by detecting candidate therapeutic targets such as cardiomyopathy and determining the related genes [Theodoris *et al*.,2022]. Fitness landscape inference is developed to map protein fitness under given environments and navigate their residue mutation effect in evolutionary trajectories [Xu *et al*., 2023]. Trajectory inference also known as pseudo-time analysis predicts the order or "progress" ranging from the original to the end cell state for single cells from genome-wide omics data [Saelens *et al*., 2019]. Noticeably, cell ordering, topology, stability, and usability of trajectory inference methods highly depend on the dimension of the dataset and the topology of the trajectory.

Domain exploration leverages biomedical text, images, video, etc., for biological domain-specific analysis such as name entity recognition, medical image extraction, medical complementary [Lee *et al*., 2020], etc. Prevalent text processing techniques of natural language processing (NLP) make numerous efforts to push the progress of mining biomedical text for name entity recognition, relation extraction, sentence similarity, document classification, natural language inference, evidence-based medical information extraction, abstractive summarization, question answering, multiple-choice question answering, etc. Analyzing terms and expressions in the biological domain corpus is pivotal for these tasks. For instance, relation extraction on PubMed enables the discovery of chemical-protein interactions where the majority of relation instances consist of single sentences. In medical vision, they specialize in visual recognition, image captioning, and medical image segmentation. Other domain-specific analyses focus on the medical complementary and alternative, for instance, grounded radiology reports, bedside decision support, augmented procedures, etc.

Multimodal integration deciphers manifold biological understanding across data modalities such as cross-modal retrieval, and multi-modal understanding. Besides the aforementioned downstream analysis tasks, many other tasks are not listed or remain to be further studied employing FMs, such as chemical-genetic interaction prediction and other modality-relevant tasks for future biological problems.

**FMs for biological domain exploration.** Modeling in domain knowledge has long been explored by foundation models in the area of natural language processing and computer vision [Liu *et al*., 2023, Kaddour *et al*., 2023]. A series of methods such as BERT [Devlin *et al*., 2019], K-BERT [Liu *et al*., 2020], GPT 3 [Brown *et al*., 2020], Dragon [Yasunaga *et al*., 2022] etc., utilized FMs to map text, images, knowledge graphs, or their combined data such as Wikidata [Denny *et al*., 2014], BoogCorputs [Zhu *et al*., 2016], and ConceptNet [Robyn *et al*., 2017] to curate comprehensive language and their complementary domain representation information. Along this line, biomedical text, images, and knowledge graphs/networks could be analyzed in the same way by transforming the text, image, or graph domain into the biological domains to solve domain shift biological problems.

Interpreting the semantic-level biological information encoded within biological text for DNA, RNA, and proteins has been a pivotal objective for FMs in biological domain exploration. BioBERT [Lee *et al*., 2020], Med-PaLM [Singhal *et al*., 2023], BioBLECTRA [Raj Kanakarajan *et al*., 2021], BLURB [Gu *et al*., 2021], BioBART [Yuan *et al*., 2022], etc., enable to shift the general domain into the biological domain by efficient tokenizer learned from unsupervised pretraining, thereby decoding potential biological operations and functions. BioBERT identifies a multitude of domain-specific proper nouns in biomedical texts leveraging its final layer representations to compute token-level BIO2 probabilities exclusively. It also employs sentence classification via a single output layer using BERT's [CLS] token representation for relation extraction (RE) and SQuAD the same architecture as BERT [Rajpurkar *et al*., 2016] for the question-answering (QA) task. With minimal architectural modifications, BioBERT accomplishes these tasks by pre-training BERT on large-scale biomedical corpora, including PubMed [Fiorini *et al*., 2018] which contains terms and expressions not included in general domain corpus and performs better in biomedical text mining. Similarly, equipping an FM with a prompt tuning that navigates it towards yielding a desired outcome is essential in the exploration of the biological domain. This procedure involves few-shot or zero-shot learning on various biological datasets. Med-PaLM combines seven professional medical question-answering datasets (MMLU clinical topics, LiveQA, MedicationQA, MedQA, MedMCQA, PubMedQA, HealthSearchQA) for aligning the model to new domains using a few exemplars. BioBLECTRA pre-trained on PubMed and PMC full-text articles introduces a replaced token prediction pretraining task with a generator and discriminator network. BLURB pretrains a biomedical language model from scratch on unannotated biomedicine text for a wide range of biomedical NLP tasks, eliminating the need for complex tagging schemes. Lastly, BioBART is a bidirectional and auto-regressive generative language model designed specifically for biomedical natural language generation tasks, complete with corresponding data. After exploring the biological domain for biomedical tasks such as name recognition and relation extraction, etc., downstream tasks in biological domain exploration such as biomedical question answering can then be achieved together via task-specific fine-tuning in BioBERT, or parameter-efficient approaches in Med-PaLM.

Besides these biological text-based domain-shift explorations, FMs also incorporate multiple modalities of data, such as image, graph, and video samples with both genes- and cell-level biological insights to improve the representation ability in various downstream tasks. Here are semi-supervised learning models CoCa [Yu *et al*., 2022], and unsupervised learning models such as GMAI [Moor *et al*., 2023] and MSA [Wu *et al*.,2023]. CoCa builds an image encoder and unimodal text decoder for multimodal pre-training on biological images and text respectively. A pre-trained CoCa model can be utilized for video action recognition tasks using individually processing multiple frames of a video through the shared image encoder. MSA breakthroughs the lack of training data without inferior performance in medical image segmentation by a medical-specific domain knowledge integrated adaptation technique, which fine-tunes only 5% around the parameters of the fundamental model for various downstream tasks. GMAI is easy to adapt to new tasks due to the acceptance of inputs and production of outputs with varying combinations of data modalities (including biomedical text, graph, and video). With minimal or no task-specific annotated data, GMAI can perform a wide array of tasks, including constructing a comprehensive perspective of a patient's health status by integrating various modalities, from unstructured symptom descriptions to continuous glucose monitor readings and patient-supplied medication logs. Downstream tasks related to visual, vision-language, and multimodal understanding could be perfectly accomplished by zero-shot transfer, frozen feature evaluation, or fine-tuning based on one of the two modules or their combination, i.e. target-specific fine-tuning.

In the context of a biological domain-shift, pre-trained FMs exhibit competitive efficacy in biological explorations tasks, such as biomedical text mining (including named entity recognition, relation extraction, and question answering), PICO (Participants, Interventions, Comparisons and Outcomes entities) extraction, and vision-language extraction, comparable to those of general domain FMs employed in natural language processing and computer vision. For example, BioBERT [Lee *et al*., 2020] (pre-trained on biological PubMed abstracts of 4.5 billion words and PubMedd Central full-text articles of 13.5 billion words) outperforms general domain foundation model BERT [Devlin *et al*., 2019] (pre-trained on Wikipedia text of 2.5 billion words and BooksCorpus of 0.8 billion words) on various biomedical downstream tasks such as biomedical question answering task improved by 12.24% MRR, biomedical relation extraction improved by 2.80% F1 score, and entity recognition improved by 0.62% F1 score. BioBLECTRA pre-trained from scratch on PubMed abstracts and PubMedBERT achieves the superior performance of mean test results on all datasets in BLURB, which is finetuned on six biomedical text mining tasks (NER, PICO, Relation Extraction, Sentence Similarity, Document Classification, Question Answering). For publicly available NCBI-Disease, it acquires 89.38% mean test results (evaluation metrics F1 entity-level for NER, Macro F1 word-level for PICO, Macro F1 for Relation Extraction and Document Classification, Pearson for Sentence Similarity, Accuracy for the rest) than the Base 88.2%. Besides pretraining mechanisms, finetuning is another key strategy in biological FMs. Compared with the ImageNet classification accuracy of CoCa (90.6%) when frozen the parameters, the performance of CoCa with the finetuning strategy is increased to 91.0 % which is higher than other image-text FMs including ALIGN (88.6%) [Jia *et al*., 2021], Florence (90.1%) [Yuan *et al*., 2021], and MetaPseudoLabels (90.2%) [Pham *et al*., 2021].

Model capacity is pivotal to biological domain exploration as well. Medical question answering accuracy on MedQA (questions about US medical licensing exam) of MedPaLM is 67.6% with 540 billion model parameters surpasses that of state-of-the-art (SOTA) FM methods PubMedBERT (38.1%, 100M) [Gu *et al*., 2021], DRAGON (47.5%, 360M) [Yasunaga *et al*., 2022], and PubMed GPT (50.3%, 2.7B) [Bolton *et al*., 2022]. MSA fine-tuned part of the model achieves the best results compared with SOTA segmentation methods with the average Dice Score of 0.893 and 0.883 on datasets AMOS and BTCV respectively. BioBART obtains competitive performance on biomedical summarization datasets exceeding BART large for 1.93/1.31/2.1 on Rouge-1/2/L on MeQSum. However, there is still a domain-shifting problem when pre-trained on biomedical scientific articles of PubMed for BioBART. Noticeably, the lack of a standard dataset for training and different training splits could result in lower scores. Additionally, the large scale of the model can also bring technical obstacles.

**FMs for biological sequence analysis.** Biological sequence analysis is one of the most important research directions in biology, handling exponentially growing sequence data related to genes, mutations, and various biological phenomena to forecast promoter regions, enhancer regions, cis-regulatory elements, splice sites, and transcription factor binding sites, among other downstream tasks related to biological sequences. Traditional models typically train identifiers using handcrafted features, necessitating an extra step of manual feature extraction. In contrast, recent works can tackle specialized tasks such as scoring variant influences, predicting gene expression, and even unseen tasks from unknown sequences leveraging implicit medical knowledge from foundation models. They provide superior prediction results of various tasks with the constraints of correlated biological theory such as the rationale of the identification of the genomic variants is that true transcription factor binding sites are more likely located with other transcription factor binding sites.

Genome-wide association studies (GWAS) have been instrumental in providing crucial biological understanding across a multitude of species. And deciphering the language of non-coding DNA to understand how DNA sequence encodes phenotypes is a major problem for the next phase of genome biology research. Supervised foundation model Enformer [Avsec *et al.*, 2021] improves gene expression prediction accuracy, noncoding variant effect prediction, and candidate enhancer prioritization from DNA sequence through integrating long-range interactions in a larger receptive field. The tripartite structure of Enformer has seven convolutional blocks with pooling, eleven transformer blocks, and a final segment that includes a cropping layer and pointwise convolutions that diverge into two organism-specific network heads. Due to the existence of polysemy and distant semantic relationships of non-coding DNA especially in data-scarce scenarios, gene regulatory code is highly complex. DNABERT [Ji *et al.*, 2021] pre-trained model for proximal promoter region identification, and subsequently fine-tuned two models, DNABERT-Prom-300 and DNABERT-Prom-scan, using TATA and non-TATA human promoters of 10,000 base pairs in length from the Eukaryotic Promoter Database (EPDnew). It pre-trained bidirectional encoder representation to capture a global and transferrable understanding of DNA sequences after fine-tuning small task-specific annotated data to visualize semantic relationships. When dealing with sequences that extend beyond 512 in length, DNABERT segments them into manageable parts and combines their representations to yield the final composite representation. To further improve its efficiency, DNABERT-2 [Zhou *et al.*, 2023] presents enhancements including a skilled tokenizer and strategies to handle input length limitations, thereby optimizing time and memory consumption while boosting model capabilities. Specifically, it treats DNA sequences as sentences and k-mer nucleotides as words, and substitutes k-mer tokenization with a statistics-based data compression algorithm noted byte pair encoding (BPE). This strategic modification enables it to establish a state-of-the-art model for multi-species genome classification that operates with $21\times$ fewer parameters and requires approximately $56\times$ less GPU time. When extracting semantic-level genome representations, existing processes tend to rely on manual design and generate unsatisfactory representations instead of refined ones which demand costly database explorations. In response to solve this problem, CLAPE-DB [Liu *et al.*, 2023] leverages pre-training and contrastive learning on vast unannotated data in an unsupervised manner with the ability to handle imbalanced data. After pretraining on the Hear Atlas ECs32, Geneformer [Theodoris *et al.*, 2022] enables the separation of N1 downstream targets and non-targets without any perturbation data. By gene dimension self-attention mechanisms, scGPT [Cui et al., 2023] can encode intricate interactions between perturbed genes and others to overcome the experimental infeasibility in the vast potential gene perturbation space.

The universal genetic code illuminates the translation of DNA into proteins, a process primarily governed by the vast information contained within the genome rather than mere sequential order. Self-supervised learning foundation method HyenaDNA [Nguyen *et al.*, 2023] leverages genome sequences across various data lengths and model sizes to overcome this problem. Pre-trained on the human reference genome it can handle context lengths of up to 1 million tokens at the single nucleotide level, representing an increase of up to 500 times over previous dense attention-based models. Protein sequences across large protein families could be generated through language models. They can enhance the performance of protein sequence downstream tasks such as predicting protein stability, detecting remote homology, and forecasting secondary structure. Nucleotide Transformer [Dalla-Torre *et al.*, 2023] incorporates information from 3,202 diverse human genomes and 850 genomes from a broad spectrum of species, encompassing both model and non-model organisms. It shows that increased diversity enhances performance compared with increased model size.

The synthesis of proteins presents vast application possibilities in biological areas such as pharmaceutical design and protein engineering. ProGen [Madani *et al.*, 2023] succeeded in generating a million artificial sequences after the fine-tuning process using the curated lysozyme dataset. It pre-trains a protein language model on 280 million raw protein sequences with additional control tags specifying protein properties to generate artificial proteins across multiple families and functions. Interactions between proteins and DNA are pivotal to vital biological processes such as replication, transcription, and splicing. xTrimoPGLM [Chen *et al.*, 2023] pre-trains a transformer framework with 100 billion parameters to address protein understanding and generation tasks with joint optimization of the two types of objectives. This approach employs a trainable multilayer perceptron (MLP) as a probe to scrutinize pre-trained representations, providing an efficient means to discern the type of protein information. Notably, during the probing phase, the parameters of pre-trained PLMs remain the same, while training solely on the MLP. Systematic prioritization obtained from sequence-based CNN instead of the binary outcome can accurately predict the TF binding intensities and measure the impact of non-coding variants on transcription factors. Furthermore, ProteinBERT [Brandes *et al.*, 2022] enables meticulous fine-tuning across an extensive spectrum of protein-related tasks in a remarkably short span of minutes. Impressively, it demonstrated results on stability prediction that were closely aligned with the pinnacle of contemporary research. ProtGPT2 [Ferruz *et al.*, 2022] generates sequences with prevalent disorders across datasets displaying 48.64%, 39.70, and 11.66% alpha-helical, beta-sheet, and coil contents, which is comparable to the natural space with the 45.19%, 41.87%, and 12.93%.

The accurate identification of splice sites is pivotal for guaranteeing precise protein translation. Among these endeavors, DNABERT outperforms SliceFinder [Wang *et al.*, 2019] on benchmark data with superior performance 0.923 of multiclass accuracy, 0.919 F1 score, and 0.871 MCC than SliceFinder which reported an accuracy of 0.833, an F1 score of 0.828, and an MCC of 0.724. Nucleotide Transformer was noted in the prediction of splice sites, where the disparity between the top-performing probing and fine-tuned models was approximately 20%. In functional variants prediction, ProGen aligned more accurately with the experimentally measured assay data from protein datasets CM and MDH with an AUC of 0.94 than the sequence generation methods from the studies that were specifically designed for these families such as ProteinGan [Repecka *et al.*, 2021] with an AUC of 0.87. Proper functional scores could help the classification of disease-related non-coding variants. In comparison to the recently published GenomicBenchmarks [Gresova *et al.*, 2022], which encompasses eight regulatory element prediction datasets, HyenaDNA establishes a new state-of-the-art across all datasets. Notably, it surpasses previous benchmarks by substantial margins, achieving an improvement of up to 20 percentage points in the task of human enhancer identification.

Protein sequences, akin to natural languages, are comprehensive information repositories, encapsulating structure and function in their amino acid sequence with unparalleled efficiency, and the pretrained foundation model can be adapted for accurately predicting their structure. ProtGPT datasets exhibit a comparable distribution of ordered and disordered regions across the two datasets IUPred3 and ordered content. Notably, the proportion of ordered amino acids in the ProtGPT2 and natural datasets are 79.71% and 82.59%, respectively, underscoring the similarity in their composition. Specifically, foundation model xTrimoPGLM achieved a 0.961 TM score in predicting VH and VL structure in antibodies, which is higher than AlphaFold2 (0.951) and other advanced methods including OmegaFold [Wu *et al.*, 2022] (0.946), ESMFold [Lin *et al.*, 2023] (0.943), IgFold [Ruffolo *et al.*, 2023] (0.945) and xTrimoAbFold [Wang *et al.*, 2022] (0.958). While we do not foresee FMs generating an entirely different distribution or domain (such as inventing a new fold that triggers an unnatural reaction), they do have the capability to considerably expand the variety of sequences sampled by evolution, thereby enhancing model performance.

**FMs for biological structure construction.** Comprehending biological secondary and 3D structures is crucial for medical treatments, such as vaccine development through the determination of mRNA structure. The task of predicting these structures poses a significant challenge for biologists, necessitating concerted efforts to improve our understanding of biological folding rules and enhance the precision of structure prediction models. Traditional biological structure construction depends on physics-based methods such as cryogenic electron microscopy, thermodynamic methods helped by experimentally measured thermodynamic parameters, and alignment-based methods [Sato *et al.*, 2021, Ding *et al.*, 2023]. Due to the high costs of wet lab experiments and the structural instability of genes like RNA, painstaking efforts have been surged in the development of computational methods. Recent foundational models make great breakthroughs in biological structure construction. They enable the establishment of a learned, RNA/protein-specific neural network to predict secondary structure accurately from its sequence and refine the predicted structure by the manipulation of tokens, the embedding of position. In practice, modeling protein structure hinges on the integration of representational data with annotations in tasks. For 3D structures, more strict constraints and implementation of pre-training tasks contribute to obtaining precise structures. In this case, a higher dimensional representation of the distance and interactions between tertiary inter-nucleotide pairs might be required.

Reconstruction of structures from sequence data is a major challenge by their labor-intensive and time-consuming characteristics. This challenge is amplified by the limited biological structure datasets [Skinnider *et al.*, 2020]. To overcome this problem, AlphaFold [Senior *et al.*, 2020], a supervised deep learning method, obtains the distance map and torsion distribution between pairs of residues from protein sequences for an efficient protein structure prediction. Foundation model AlphaFold2 [Jumper *et al.*, 2021] further improves the accuracy with a certain noisy student self-distillation approach, generates a new dataset of predicted structures, and predicts the structure of diverse sequences from Uniclust30. Without directly using structure data in PDB datasets, they jointly employ multiple sequence alignments (MSAs) and structure features to obtain final constructions. ProteinBERT [Brandes *et al.*, 2022] recovers uncorrupted data from the corrupted inputs by randomly replacing tokens and adding random false annotations to force the model to predict annotations from sequence alone. It obtains superior performance covering diverse protein properties including protein structure, post-translational modifications, and biophysical attributes. ProGen [Madani et al., 2023] utilizes a protein language model trained on millions of raw protein sequences to generate artificial proteins with a structural divergence that is conducive to predicting protein secondary structure.

Compared with traditional linear regression methods shaped to hidden representations using annotated data to predict the secondary structure, or shaped two separate linear projections of sequence position pairs for tertiary structure, ESM-1b [Rives *et al.*, 2021], an unsupervised foundation model, trains a deep contextual language model on 86 billion amino acids across 250 million protein sequences, which enables a scale combination in data and model capacity. NetSurf supplants the conventional logistic regression linear layer with a deep neural network in predicting secondary structures [Klausen *et al.*, 2019]. xTrimoPGLM unravels secondary structures of proteins depending on the classification task on helices, strands, and various turns like coils. CLAPE-DB [Yufan *et al.*, 2023] combines the pre-trained protein language model ProtBert [Elnaggar *et al.*, 2021] and constructive learning that discovers a representation space to predict ligand-binding sites of a protein sequence. HyenaDNA adds gradient checkpointing to predict chromatin profiles including transcription factor binding profiles, DNase I-hypersensitive sites, and histone marks, which reduces the memory footprint by 3x.

For tertiary structure, we can extract a binary contact map from the hidden representations of sequences, which offers an alternative to the conventional method that applies two distinct linear projections to hidden representations. FMs, by contrast, can directly predict 3D structures and positions of biological targets. ProtGPT2 [Ferruz *et al.,* 2022] was found to generate protein sequences that not only mirror the amino acid and disorder characteristics of natural proteins but also carve out a unique niche within the existing protein landscape. Uni-MoI [Zhou et al., 2023] predict 3D positions for various downstream tasks like molecular property prediction and molecular conformation generation by two pre-trained models: a molecular model, pre-trained using 209 million molecular conformations, and a pocket model, pre-trained with 3 million candidate protein pocket data. According to the stationary-action principle [Richard *et al*., 2017], it completes self-supervised pre-training on selected positions with minimal delta positions from random positions instead of using a masking strategy to recover the correct 3D position.

Rapid prediction of protein structure is indispensable for protein design and examination of allelic variation or disease mutations. RGN2 [Chowdhury *et al*., 2022] achieves a remarkable reduction in computation time by up to 106-fold outperforming AlphaFold2 in the analysis of orphan proteins and various classes of designed proteins. It utilizes the Frenet-Serret formulas to embed a reference frame at each Cα carbon, then the backbone can be easily constructed by a series of transformations, i.e., a protein language model AminoBERT. Emerging FMs have been proposed to learn the distinctive representations of non-coding RNAs, which align with downstream secondary/3D structure prediction, SARS-CoV-2 genome structure and evolution prediction, protein-RNA binding preference modeling, and gene expression regulation modeling. RNA-FM [Chen *et al*., 2022] adopts self-supervised learning taking advantage of 23 million non-coding RNA sequences to infer their sequential and evolutionary structural information on a large amount of unannotated data for higher generalizability and performance. It leverages four Evoformers as its foundational structure, and further stacks on top to an Equivariant Graph Neural Network (EGNN), serving as a predictor of 3D atomic coordinates. Noticeable, pre-training mechanisms in recent works play a crucial role in structure construction. Guo *et al.* [Guo *et al*., 2022] propose a self-supervised pre-training model to learn hierarchical structure embeddings from protein tertiary structures to improve predicting efficiency. Moreover, McDermott *et al.* [McDermott *et al*., 2023] impose relational structure constraints on the pre-training framework and take a pre-training graph as an auxiliary input, whose performance is supported by theoretical results.

Foundation models in the construction of biological structures significantly transcend the boundaries of traditional structure prediction. RNA-FM achieves 94.1% and 70.4% of F1 score on ArchieveII600 and bpRNATS0 respectively, which surpasses SPORT-RNA [Singh *et al*., 2019] by 22.8% and 7.5% and is notably higher than the SOTA UFold [Fu *et al*., 2022] by 3.4% and 4.0% respectively. RGN2 exhibits superior performance over other methods when applied to proteins that are rich in single helices and bends, or those that feature hydrogen-bonded turns interspersed with helices. ESM-1b indicates that incorporating features obtained by the transformer results in an absolute accuracy improvement of 0.9% and 2.5% respectively compared with the HMM profiles used by NetSurf [Klausen *et al*., 2019] on the CB513 test set for secondary structure prediction. This suggests that transformer features provide information that is not captured by the MSA-derived features. Moreover, the integration of supplementary information can provide practical solutions that are essential for enhancing our comprehension of biological structure. For example, AlphaFold combines bioinformatics and physical approaches to build components from PDB data, which enables handling mission physical context in challenging cases like intertwined homomers.

As for 3D pose prediction of protein-ligand complexes, Uni-Mol predicts 80.35% of binding poses with an RMSD less than or equal to 2Å, better than popular docking methods such as Vinardo (62.81%), Autodock Vina (64.56%), and Smina (65.26%). On the RNAcontack Test80 dataset, 3D closeness prediction of RNA-FM achieves a superior of 0.88 in the Top-10 long-range top precision. This performance is significantly superior compared to other methods which typically fall within a range of 0.48 to 0.68. When dealing with small-scale data tasks, RNA-FM confirms that the transfer learning employing pre-trained parameters of ResNet32 on bpRNA-1m enables improvement of the task performance by another 20 points than simple ResNet32 with RNA-FM. Furthermore, when dealing with large-scale data tasks, fine-tuning RNA-FM together with downstream modules enables higher performance. RNA-FM 3D distance prediction attains a PMCC of 0.8313 when combining sequence encoding, MSA covariances, and RNA-FM embeddings higher than that of combining sequence encoding, MSA covariances, and secondary structures (0.8218 PMCC).

The potential of FMs extend beyond structure prediction, offering practical solutions for tasks integral to their application in gene expression, such as the identification of binding sites for transcription factor proteins. CLAPE-DB demonstrates superior performance with AUC values of 0.871 and 0.881 on two benchmark datasets, TE46 and TE129 in DNA-binding sites prediction. It outperforms the latest advanced sequence-based models, including DNAPred [Zhu *et al*., 2019] with AUC values of 0.845 and 0.730, NCBRPred [Zhang *et al*., 2021] with AUC values of 0.823 and 0.713, and SVMnuc [Su *et al*., 2019] with AUC values of 0.812 and 0.715. More types can be further identified with increasing genomic profiling data and 3D genome contact maps.

**FMs for biological function prediction.** Biological functions maintained a high attention towards intronic regions of genes contribute greatly to complicated disease understanding. Traditional function prediction models mainly classify targets into one or more categories of collected function datasets such as Gene Ontology (GO) [Ashburner *et al*., 2000] that describes function by hierarchical ontologies including molecular functional (MF), biological process (BP), and cellular component (CC) [Mi *et al*., 2019, Gligorijević *et al*., 2021, Kulmanov *et al*., 2022]. Although GO has more than 50,000 classes, existent function taxonomy is immature, incomplete, and imbalanced, hence remain challenges in predicting them correctly in a complex large space even assisted with biological features. Furthermore, as highly variable genes (HVGs) are mainly selected from the expression variance across the entire dataset, there is a potential of missing crucial genes of less common cell types. Here, biological function prediction FMs could break the dilemma and be resilient to data noise and variability.

For example, Geneformer [Christina *et al.*, 2023], a context-aware, attention-based deep learning model pre-trained on a corpus of 29.9 million transcriptomes, can accurately predict disease genes and their targets, and can be fine-tuned for a variety of downstream tasks related to chromatin and network dynamics using only a small number of task-specific training examples. All genes are reorganized based on their gene expression and fed into the transformer for training, in which pre-trained model could be commonly applied to cardiomyopathy disease modeling, pinpointed therapeutic targets and their experimental suppression led to significant improvement in cardiomyocyte contraction in an iPSC-based model. The attention heads are learned in an unsupervised manner for distinct class prediction without previous biological function knowledge of any gene. Indeed, biological knowledge graphs and networks provide a robust foundation for relational learning models such as interlinked identifiers, thereby enhancing our understanding of complex biological systems [Walsh *et al.*, 2020]. Altered data can also be employed as a semi-supervised way to generate detailed representations for function predictions. DNABERT [Zhou *et al.*, 2023] provides an accurate prediction of functional genetic variant candidates for around 700 million short variants in dbSNP on selected variants only in high-attention regions and repeats the predictions based on altered allele sequences. xTrimoPGLM [Chen et al., 2023] provides four distinctive masking strategies to redesign the selected sequence and evaluates the implications of a synthesized protein sequence associated with a specific biological function. For instance, the task of antibiotic resistance predicts whether a protein sequence exhibits sensitivity to a particular bacterium. ProtST [Xu et al., 2023] proposes an unsupervised multimodal integration pretraining framework on both protein sequences and biomedical texts, outperforming the sequence-based model ESM-1b [Rives et al., 2021] on protein function annotation. RNA-FM [Chen *et al.*, 2022] leveraging embeddings pre-trained on ncRNAs to model the function of the 5' untranslated region in mRNA showing its ability to handle non-coding sequences.

Existing methods necessitate the pre-processing of raw data through the selection or manipulation of genes, such as HVG selection, manual selection of marker genes, and PCA. This is primarily due to their limited ability to efficiently model high-dimensional data. To overcome this challenge, scBERT [Yang *et al.*, 2022] pre-trained on large-scale unannotated scRNA-seq data in a self-supervised learning manner to overcome the batch effect, enlarge sequence length, and improve the model's generalizability employing Performer [Choromanski *et al.*, 2020], a matrix decomposition transformer. Foundation model scGPT [Cui *et al.*, 2023] employs an in-memory data structure tailored specifically for non-sequential omic data, enabling the storage of hundreds of datasets and facilitating rapid access to manage large-scale data. It pre-trains over 10 million cells stored by the in-memory data structure and converts all expression counts into relative values by a novel value binning technique. It also offers reusable fine-tuning pipelines and objectives, specifically designed for a range of downstream tasks including cell type annotation, gene network inference, multiomic integration, and perturbation prediction, facilitating users to effortlessly apply the pre-trained model. While pre-training on extensive unannotated data can yield transferable knowledge for downstream tasks, it's crucial to acknowledge the divergence that exists between the pre-training and fine-tuning stages. Specifically, the optimization objectives of downstream tasks could be shifted according to the task. L2P-GNN [Lu *et al.*, 2021] employs a dual adaptation mechanism at both node and graph levels to encode local and global information, thereby enabling the optimization objectives of downstream tasks to be tailored to the task at hand and facilitating biological function prediction using 88,000 annotated subgraphs for a 40-binary classification task.

Another limitation of available training data is the imbalance and extremely highly similar subtypes result in the inferior performance of the SOTA method. For example, the cell type annotation on the Zheng68k peripheral blood mononuclear cell (PBMCs) dataset could not achieve an accuracy surpassing 0.71. By contrast, FM scBERT acquires 0.759 on cell type annotation in the same situation. Meanwhile, scBERT enables capturing long-range interactions and achieves higher performance on both known and unknown classes whose accuracy of unknown class is 0.329 compared with that of SciBet [Li *et al.*, 2020] (0.174) and scmap_cluster [Kiselev *et al.*, 2018] (0.174) and known class prediction accuracy is 0.942 compared with that of SciBet (0.784) and scmap_cluster (0.666). The rank value encoding sometimes provides a straightforward way to understand the transcriptome, i.e., gene expression, which helps the model focus on genes that are important in distinguishing different cell states. Genomeformer mapping the network hierarchy with large amounts of transcriptional perturbation data significantly boosts the prediction of central versus peripheral factors (AUC 0.81) compared to other methods (AUC 0.59-0.69). Significantly, data quality will be mirrored in the performance of individual input networks. Although the functional tasks, which are tied to the distribution of pre-trained data, have not yet surpassed structural tasks in terms of improvements, biological function prediction FMs outperform baselines by a large margin and acquire higher accuracy.

**FMs for multimodal integration biological problems.** Recently, language understanding research indicate that the text corpora pre-trained model is surprisingly effective at the task of image synthesis, which brings a perspective on multimodal analysis [Saharia *et al.*, 2022]. Traditional biological models mainly focus on unimodal information, and have difficulties in handling multimodal data or multi-level data [Cao *et al*., 2022, Ciciani *et al*., 2022, Xu *et al*., 2023]. While foundation models enable a more general understanding of targets by semi-supervised and self-supervised pre-training, for example, they are also vulnerable to perturbation and need a specific fine-tuning approach. Multimodal integration enables a deeper understanding of diverse topics, assisting in leveraging medical data for accurate diagnosis, treatment recommendations, and medical research support. Thus, it holds significance for multimodal integration in histology diagnostic imaging and genomics molecular profile data, etc., for a systematic study of biological samples [Chen *et al*., 2022]. CoCa [Yu *et al.* 2022] introduces a contrastive loss on image-text embeddings and a captioning loss on multimodal decoder outputs for image-text involving tasks such as crossmodal retrieval and multimodal understanding. scGPT [Cui *et al*., 2023] enables multi-omics prediction through generative AI with the integration of expression and new condition tokens to extend the embedding architecture to multiple modalities, batches, and states. Specifically, protein sequences together with their textual property descriptions achieve the end goal of protein representation learning: protein function acquisition. ProtST [Xu et al., 2023] improves the original representational capacity of the protein language model with protein data of varying granularities and properties through the application of multimodal representation alignment and multimodal mask prediction.

In the future, it is important for biological foundation models to incorporate more diversity data, temporal data, perturbation data, etc. For example, KGs analysis suffers from low accuracy due to the data unbalance problem in that the richer entities own more relations and information but the scarce ones will not be fully represented with limited information. A possible solution, even if it would possibly raise prices on model designing and training, would be to incorporate varying forms of biological data such as sequence data, structure data, chemical compounds, etc. With the appropriate use of exponentially growing biological data and advanced foundation models, both clinically and biologically expected outcomes could be obtained.

## Challenges and Opportunities

Despite the remarkable progress made in bioinformatics, foundation models still face some challenges in solving biological problems. Biological data containing complex information in living organisms represented with various data scales, different styles, and multiple data types bring a lot of challenges to FMs [Ruiz *et al.*, 2021]. Employing deep learning modules in FMs is also a double-edged sword [Eraslan *et al.*, 2019]. It makes large biological data analysis possible but requires a great deal of computing resources, conceives a massive of model parameters, and has low explainability and reliability. These challenges and potential opportunities for promising biological areas illustrated in Fig. 3 are presented separately as follows.

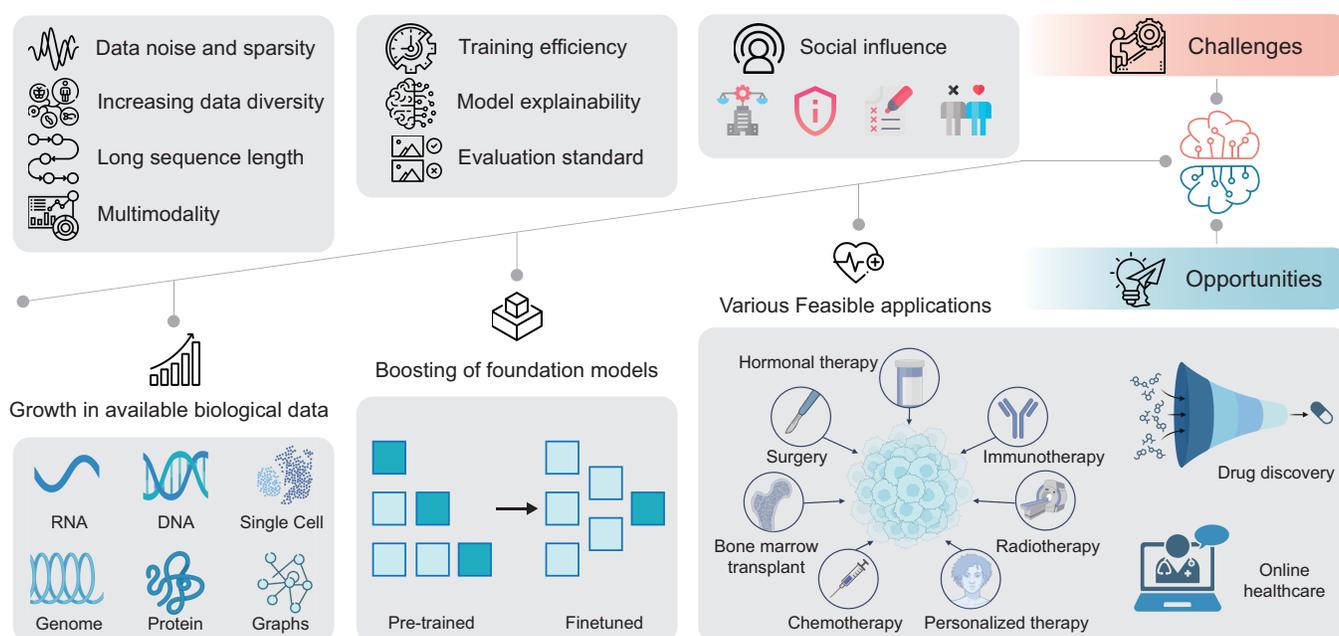

**Figure. 3 Challenges and Opportunities in Applying Foundation Models for Biological Problems.** FMs for solving biological problems have been met with challenges related to biological data, model structures, and their social influence, which in return catalyzes opportunities in bioinformatics due to the increasing availability of biological data, the enhancement of foundation models, and their diverse real-world applications. From the top of Fig. 3, the challenges encompass data noise and sparsity, increasing data diversity, long sequence length, and multimodality in biological data collecting; training efficiency, model explainability, and evaluation standards in model design and construction; and social influences such as ethics and fairness, privacy restriction, potential misuse, and social bias. Conversely, at the bottom of Fig. 3, opportunities are emerging with the growth of biological data types and volumes, including RNA, DNA, single cell genomics, protein, and knowledge graphs/networks; the enhancement of foundation models, particularly pre-trained mechanisms; and a wide range of applications spanning surgery, hormonal therapy, immunotherapy, radiotherapy, personalized therapy, chemotherapy, bone marrow transplant, drug discovery, and online healthcare. These developments signal a promising trajectory for the application of FMs in bioinformatics.

**Challenges**

**Data noise and sparsity.** One of the key goals of FMs is to extract embeddings for downstream analysis or support other biological problems [Poli *et al.*, 2023, Jeliazkov *et al.*, 2023]. However, it is still challenging for them to tackle sparse data or corrupted noise data. The sparsity of biological data is mainly caused by data collection deficiencies, e.g., data captured under different chemistry versions and depths, and unbalanced research concentrating on popular phenomena and generating entities that are already rich. Noises and biases usually appear in different selection strategies, experiment conditions, etc. These entities are unavailable for biological downstream tasks, especially relational discovery tasks. For example, there will be an indirect data leakage when a relational model designed to extract information about drug-drug combination interactions wants to take drug-protein interactions as the extra supporting data, even when they are targeting the same protein, i.e. they are truly related [Walsh *et al.*, 2020]. Although some of the FMs indicate such an issue could be relieved with a careful review of these data and deep investigation of the phenomenon, they are still vulnerable to data corruption that exists in the current evaluation sets or future capture scenarios in the real world. Therefore, we suggest incorporating different forms of biological data to make up for the unexplored data, enhance the model representation ability, and understand the entities that are under-represented.

**Increasing data diversity.** Another obstacle in bioinformatics for FMs is the increasing data diversity in evolution. On the one hand, increasing data diversity has the potential to improve model performance efficiently without the need to increase model capacity [Rives *et al.*, 2021]. On the other hand, diversity with task-unrelated data in a real-world situation can hardly be transferred to downstream tasks [Wang *et al.*, 2019, Wang *et al.*, 2022]. Along this line, if foundation models could generate new sequences that are biologically active or pseudo-samples of previous classes and transfer their knowledge to design new proteins with different functions, they could both condense the biological statistics and optimize the biological problem-solving ability for special tasks. Improving the robustness might be another direction to help FMs overcome the barriers that hinder the implementation of increasing diversity. Although simply adding bias to FMs may improve their performance, their capacity to exploit deeper and broader features is also limited. Therefore, we could provide higher capacity FMs for increasing data diversity to perform better in bioinformatics.

**Long sequence length.** Biological sequence provides great potential in solving various biological problems, but the long sequence length raises huge challenges in training models. There are around 3 billion nucleotides in a single human, 5 million in the bacterial, and 30 thousand in the virus, these long sequence lengths bring an extreme gradient variance thereby improving the instability and reducing the efficiency when training a model. The essential reason for the correlation between long sequences and training instability has not been completely deciphered. Motivated by Sortformer [Press *et al.*, 2020], Li *et al.* [Li *et al.*, 2022] attempt to enable stable training with reduced cost by improving data efficiency. They try to overcome the stability-efficiency dilemma through a sequence length warmup method which trains the model from short sequences to longer length sequences gradually with larger training batch sizes and learning rates. However, the varying lengths of sequences implemented in this method are directly obtained by truncation and sacrifice the information of dropped data. Causal relationships, prerequisites, or other significant factors have not been sufficiently represented and certified, which could be further improved by leveraging data localization, structure, function, or other chemical and biological rules and relations. Thereby, longer sequence lengths and higher dimensional gene maps could be easily adopted in analyzing biological problems.

**Multimodality.** Experimentally, models with higher capacity yield better understanding and representation. However, it is not the decisive factor influencing the performance in downstream task analysis. In bioinformatics, there are multiple types of data (text, image, multi-dimension structure, molecule, etc.) acquired from varying scale biological targets (DNA, RNA, protein, single cell genomics, etc.) and different recording technologies with different annotations, which brings the multimodality challenge for FMs. Specifically, different representations of the same gene from different inputs instead of the same representation are substantially enriched in alpha cells and suitable for the cell type annotation downstream task [Yang *et al.*, 2022]. Multimodality refers to the integration of multiple types of biological data, such as scRNA-seq, scATAC-seq, and ChIP-seq. However, the challenge lies in the uneven data availability across these modalities, with scRNA-seq having more data compared to scATAC-seq and ChIP-seq. This disparity poses significant challenges for comprehensive multimodal analysis. To generate accurate representations from diverse multimodal biological data, the FMs should pay more attention to both feature-level and semantic-level training strategies to unify the biological knowledge. For instance, despite their similar model capacity, Transformers, a deep learning structure commonly employed in FMs, exhibit superior prediction capabilities compared to LSTMs [Zhou *et al.*, 2021]. Therefore, exploiting the inherent representing strengths of FMs is significant for more diverse and complex features to tackle different tasks.

**Training efficiency.** Pre-training is a crucial step for most of the FMs to maintain coherence within each shot, which in turn affects the overall quality of a summary. However, their high computational costs on huge amounts of data remain a large barrier to their implementation (e.g., AlphaFold2 needs several weeks of training on up to 200 GPUs). To improve the efficiency in analyzing big data, previous approaches, for instance, leverage attention mechanisms such as FlashAttention [Dao *et al.*, 2022] and Multi-query attention [Ainslie *et al.*, 2023], quantization [Yao *et al.*, 2022], kernels [Hijma *et al.*, 2023], sparse activation [Xu *et al.*, 2023], and other advanced mechanisms to reduce the model's training and detection time. Similarly, substantial redundant computations could be cut down in FMs with advanced technologies in terms of removing unimportant parameters, reducing memory consumption, enhancing convergence rate, paralleling data, and fully utilizing the generative and adaptive capabilities of models [Sato *et al.*, 2021]. In general, more efforts could be made along this line to improve the efficiency of foundation models made already.

**Model explainability.** It is also challenging to provide interpretability of FMs in each step and acquisitions with logical evidence in bioinformatics. Clear and strong explainability and interpretability are significant factors of highly comparative prediction accuracy enabling a wide range of biomedical and healthcare applications to explain the model and results to consumers and researchers [Cui *et al.*, 2022]. Some efforts have been made to explain them in biological applications such as scBERT, which explains the contribution of genes and their interactions by attention weights of the self-attention mechanism in the model for gene exploration and decision-making tasks. Thereby the top genes could be visualized by the weights and analyzed in the following stages [Yang *et al.*, 2022]. However, this work relies only on structural results, which neither indicate the importance of each node nor explain the reason for reliable results obtained by the model. We envision that FMs dramatically improve interpretability and explainability by incorporating knowledge graphs and networks to narrow the gap between FMs and experts for solving more complex biological problems.

**Evaluation standard.** The design of traditional AI-based models for a specific task in computer vision or natural language processing makes it easy to evaluate the results as the model performance fits the predefined metrics. However, foundation models face various downstream tasks as well as unseen tasks, making it uniquely challenging to anticipate all of the modes and set an evaluation standard for these methods. Current qualitative evaluations mainly focus on certain modules such as Machine Reading Comprehension (MRC) within a complete QA pipeline, instead of a previously unseen task, e.g., diagnosing disease in a brain MRI [Moor *et al.*, 2023]. Moreover, the general domain evaluation takes no account of the effects of rich biological regulations such as biomedical synonymous relationships [Jin *et al.*, 2022]. To evaluate the model performance and output quality that convey model uncertainty accurately, which in turn prevents the occurrence of overly confident assertions, biological knowledge of radiology, pathology, oncology, and other specialties might be required.

**Social influence.** As foundation models allow researchers and consumers to receive help across the medical treatment and health sciences for improved medical research, human health, and ecological and social environments, serious challenges have emerged due to the adverse effects of ethics and fairness, privacy restriction, potential misuse, and social bias, etc [Weidinger *et al.*, 2021, Yu *et al.*, 2022]. For example, gathering biological data in an open environment is the key to biological research, however, data sharing presents an enormous challenge for researchers whose data lies at the center of their experiments usually restricted to privacy [Kaddour *et al.*, 2023]. Moreover, over-reliance on models harms patients and may cause disease misdiagnosis from health disparity[Vyas *et al.*, 2020, Eneanya *et al.*, 2022]. Along this line, we suggest the endeavors of FMs to build quality assessments for various tasks and utilization. Besides these challenges, social support (beneficial to society) for both biological data and FMs is an essential factor that directly affects their development process and speed. Hence, it is necessary to ensure the security of models, the privacy of patients, and the safety of both ecological and social environments, and take exploration with legal and ethical guarantees.

## Opportunities

**Biological data.** Due to the exponential growth in available biological data, e.g., for RNA secondary structure prediction there are bpRNA-1m [Danaee *et al.*, 2018] (102318 sequences from 2588 RNA families), RNAStralign [Tan *et al.*, 2017] (30451 sequences from 8 RNA families), ArchiveII [Sloma *et al.*, 2016] (3975 sequences from 10 RNA families), the performance of FMs on downstream tasks is expected to boost. Furthermore, incorporating diverse forms of biological data, such as sequence data, structure data, and chemical compounds, could potentially enhance the model's capabilities and robustness to noise and outliers. Besides simple datasets, e.g., CellxGene Single-Cell Datasets [Hyman *et al.*, 2022] in sequence analysis, and augmentation datasets, e.g., ProtDecribe [Xu *et al.*, 2023] (enhance protein sequences with textual descriptions of their functions) in function construction, there is a huge amount of data that has not fully utilized by FMs, which could bring new insights and understanding of important directions. Complex combinations of biological information or conditional data, for instance, have not yet been widely analyzed in FMs. Specifically, multibiomics data containing microorganisms and material in surroundings offers the potential to understand the information flow that is fundamental to disease processes [Hasin *et al.*, 2017]. Additional related information could also be beneficial for biological problem-solving [Fu *et al.*, 2022]. For instance, AlphaFold, a leading model in the field, combines bioinformatics and physical approaches to build components from Protein Data Bank (PDB) data. This approach enables it to handle physically challenging contexts, such as intertwined homomers. With the right application of biological data and advanced foundation models, we could achieve outcomes that meet both clinical and biological expectations. Thereby, we can enhance FMs towards a specific target, such as the structural features, and optimize the related tasks in a way that does not affect the target. After this optimization process, for example, to optimize the structural features in FMs providing information that is not captured by the MSA-derived features [Chen *et al.*, 2022], we can expect improved results.

**Foundation models architecture.** Supervised learning in traditional methods depends heavily on a large volume of labeled data and tends to have limited generation capabilities. As a result, foundation models have emerged as an alternative approach. Due to the similarity between biological data in bioinformatics and digital data in fields like natural language processing and computer vision, the application of FMs in bioinformatics become straightforward and convenient. It adopts pre-trained, and fine-tuned, few-shot, or zero-shot learning manner to obtain a comprehensive biological map or establish the model within such a diverse environment remains a challenge. Transitioning from a derivable approach to a multi-focus framework also presents difficulties. In this respect, we can discuss FMs from two perspectives. When biological data and model size are controllable, we might design different strategies for different data and tasks. On the other hand, when dealing with particularly large models that contain extensive biological information and a massive number of parameters, ensuring quick and stable learning becomes a crucial factor. Despite these feasible efforts, the current cognitive abilities of FMs still fall short of expectations. Understanding biological processes, such as predicting how proteins fold, remains a complex problem in bioinformatics. To this end, developing new training strategies for FMs is of paramount significance.

**Feasible applications.** In the field of bioinformatics, FMs provide several opportunities for disease understanding, drug discovery, online healthcare, etc. For disease understanding, especially the therapy of cancers, exploring the detailed landscape of the microenvironment from different perspectives is significant. For example, the study combined both scRNA-seq and spatial transcriptome (ST) facilitate to the analysis of complex tumor. However, effectively microenvironment (TME) in colorectal (CRC) and to understand the crosstalk between tumor-infiltrating fibroblasts and myeloid cells involved in CRC [Peng *et al.*, 2023]. Along this line, FMs could further introduce the multi-modal deep learning model exploring scRNA-seq, spatial transcriptome, bulk RNA-seq, and other information from basic experiments to provide the physiological function of targets and gradually replace the analytical ideas of building cancer prognosis models.

The obstacles faced in drug discovery mainly emerged in biological target identification bound to treat and disease, which typically requires years of extensive customization and experiments [Schneider *et al.*, 2018]. FMs that enable in-depth analysis of biological targets such as genes, proteins, and molecules and provide their sequence, structure, and function information and representations have great potential to boost this discovery process. By providing a wide search space where corresponding phenomena (e.g., polypharmacy side-effects, viral mutations) could also be discovered, they will impact therapeutic design and bring success in discovering new drugs without paying extra time and money in wet lab experiments for insurance [Huang *et al.*, 2021]. Moreover, given a patient's genetics, genome, and health history, drugs of personalized medicine could be designed by FMs. They benefit drug prediction from medical images to gene and molecular measurements across multimodal data of patients [Capobianco *et al.*, 2022]. In the following clinical trials, drug and treatment problems can be tracked by them as well from potential failures prediction and eligible patient matching, etc [Liu *et al.*, 2021, Moon *et al.*, 2023].

Traditional healthcare and biomedicine services are mainly provided directly by health workers and doctors with the help of expensive tests and equipment, which wastes a lot of resources and time in the diagnosis of rare and emergent diseases [Gamache *et al.*, 2018]. FMs, a lightweight computer mechanism enabling online health care, can be recognized as central storage in bioinformatics to provide a massive of medical knowledge for professional diagnosis, scientific therapies, and healthcare administration without the expensive consumption of medical resources. Online healthcare provides the potential for FMs to be the backbone of any healthcare system and greatly reduce the impact of urgent pandemic crises (e.g., COVID-19) [Kocher *et al.*, 2021]. Their applications, such as question-answering systems and healthcare assistive robots, can help users acquire salient medical information when dealing with obstacles they may face [Demner-Fushman *et al.*, 2020]. Their understanding of diseases can support both doctors and biological researchers to improve their efficiency and make the most of the medical data and resources [Wornow *et al.*, 2023]. As a result, online health care achieved by FMs is a promising direction for consumers, researchers, and governments.

## Conclusions

To conclude, foundation models are advanced and efficient in solving biological problems including but not limited to three core biological problems: biological sequence analysis, biological structure construction, and biological function prediction on various downstream tasks. Other problems like biological domain exploration and multimodal problems are also analyzed and compared with traditional methods. Challenges and opportunities across biological data, foundation models, and social influence to provide an efficient way to maintain the advantages of FMs while solving emerging biological problems better are attached in the end.

## Ethical Statement

There are no ethical issues.

## Acknowledgments

The work described in this paper was partially supported by a grant from the Research Grants Council of the Hong Kong Special Administrative Region, China [Project No.: CUHK 24204023], and a grant from Innovation and Technology Commission of the Hong Kong Special Administrative Region, China [Project No.: GHP/065/21SZ].

**Table 2 A summary of foundation models in bioinformatics.** The table summarizes the model categories, targets, deep module type, and technical advancement of foundation models for tackling biological problems (DE: Domain Exploration, SA: Sequence Analysis, SC: Structure Construction, FP: Function Prediction, and MP: Multimodal Problems). FMs are categorized by their pretraining paradigms: supervised learning, semi-supervised learning, and unsupervised learning. Target biological data types include DNA, RNA, protein, single cell genomics (scGenomics), biomedical text/image/video, and knowledge graph/network. Various deep modules enhance the performance or interpretability of FMs, such as MLP: multilayer perceptron, CNN: convolutional neural network, and Transformer.

| Model Name | Biological Problem | Model Category | Targets | Deep Module Type | Technical Advancement | Author Name, Publication Year |
|---|---|---|---|---|---|---|
| BioBERT | DE | Unsupervised learning | Biomedical text | Transformer | Adapt for biomedical corpora by pre-trained BERT on large-scale biomedical corpora | Lee *et al.*, 2020 |
| BioELECTRA | DE | Unsupervised learning | Biomedical text | Transformer | A biomedical domain-specific language model introducing a replaced token prediction pretraining task with generator and discriminator network | Kanakarajan *et al.*, 2021 |
| BLURB | DE | Unsupervised learning | Biomedical text | Transformer | Pretrain biomedical language model from scratch for a wide range of biomedical NLP tasks instead of using complex tagging schemes | Gu *et al.*, 2021 |
| BioBART | DE | Unsupervised learning | Biomedical text | Transformer | A bidirectional and auto-regressive generative language model for biomedical natural language generation tasks along with corresponding data | Yuan *et al.*, 2022 |
| CoCa | DE, MP | Semi-supervised learning | Biomedical text, image | Transformer | Use a contrastive loss on image-text embeddings and a captioning loss on multimodal decoder outputs for image-text involving tasks | Yu *et al.*, 2022 |
| Med-PaLM | DE | Unsupervised learning | Biomedical text | Transformer | Introduce HealthSearchQA dataset, propose a human evaluation framework, and present instruction prompt tuning for aligning LLMs to new domains using a few exemplars | Karan *et al.*, 2023 |
| MSA | DE | Unsupervised learning | Biomedical graph | MLP | A medical image segmentation model that fine-tunes the pre-trained SAM by integrating the medical specific domain knowledge | Wu *et al.*, 2023 |
| GMAI | DE | Unsupervised learning | Biomedical text, graph, video. | Transformer | Adapt to new tasks due to the acceptance of inputs and production of outputs with varying combinations of data modalities | Moor *et al.*, 2023 |
| DNABERT | SA, FP | Unsupervised learning | DNA | Transformer | Use pre-trained bidirectional encoder representation to capture a global and transferrable understanding of genomic DNA sequences | Ji *et al.*, 2021 |
| Enformer | SA | Supervised learning | DNA | Transformer | Use a larger receptive field to improve gene expression and promoter-enhancer interactions prediction | Avsec *et al.*, 2021 |
| HyenaDNA | SA, SC | Unsupervised learning | DNA | MLP \| CNN | Use a sequence length scheduling technique to stabilize training, and leverage longer context length to adapt to novel tasks | Nguyen *et al.*, 2023 |
| Nucleotide Transformer | SA | Unsupervised learning | DNA | Transformer | Build and pre-train foundational language models in genomics, across different genomic datasets and parameter sizes | Dalla-Torre *et al.*, 2023 |
| ProteinBERT | SA, SC | Unsupervised learning | Protein | Transformer | Pretrain protein language model with gene ontology annotation prediction task for both local and global representations | Brandes *et al.*, 2022 |
| DNABERT-2 | SA, FP | Unsupervised learning | DNA | Transformer | Adapt Byte Pair Encoding (BPE) to improve computational efficiency, and employ multiple strategies to overcome input length constraints | Zhou *et al.*, 2023 |

Table 2 A summary of foundation models in bioinformatics. (continued)

| Model Name | Biological Problem | Model Category | Targets | Deep Module Type | Technical Advancement | Author Name, Publication Year |
|---|---|---|---|---|---|---|
| ProtGPT2 | SA, SC | Unsupervised learning | Protein | Transformer | A generative language model trained on protein space to learn the protein language and produce sequences to sample any region | Ferruz *et al.*, 2022 |
| ProGen | SA, SC | Unsupervised learning | Protein | CNN \| Transformer | A protein language model trained on millions of raw protein sequences that generate artificial proteins across multiple families and functions | Madani *et al.*, 2023 |
| xTrimoPGLM | SA, SC, FP | Unsupervised learning | Protein | CNN \| Transformer | A pre-training framework to address protein understanding and generation tasks with joint optimization of the two types of objectives | Chen *et al.*, 2023 |
| CLAPE-DB | SA, SC | Unsupervised learning | Protein | CNN | Combines pre-trained protein language model and constructive learning to predict DNA binding residues | Liu *et al.*, 2023 |
| Geneformer | SA, FP | Unsupervised learning | scGenomics | Transformer | A context-aware, attention-based deep learning model pre-trained on a large-scale corpus and can be transferred to diverse fine-tuning tasks | Theodoris *et al.*, 2022 |
| scGPT | SA, SC, FP | Unsupervised learning | scGenomics | Transformer | A single cell foundation model through generative pre-training on over 10 million cells stored by an in-memory data structure | Cui *et al.*, 2023 |
| ESM-1b | SC, FP | Unsupervised learning | Protein | Transformer | Use unsupervised deep language model to acquire protein structure and function directly from sequences | Rives *et al.*, 2021 |
| AlphaFold2 | SC | Unsupervised learning | Protein | Transformer | Improve the AlphaFold by employing an SE(3)-equivariant transformer with an attention mechanism to represent their interactions and distances | Jumper *et al.*, 2021 |
| RGN2 | SC | Unsupervised learning | Protein | Transformer | Combine a differentiable recurrent geometric network (RGN) with a transformer-based AminoBERT protein language model to generate backbone structures from unaligned proteins before refinement | Chowdhury *et al.*, 2022 |
| Uni-Mol | SC | Unsupervised learning | Protein | Transformer | A 3D position predict model by a 3D molecular pre-training framework along with the candidate protein pre-training for various downstream tasks | Zhou *et al.*, 2023 |
| RNA-FM | SC, FP | Unsupervised learning | RNA | Transformer | Use self-supervised learning to train 23 million non-coding RNA sequences and infer their sequential and evolutionary information | Chen *et al.*, 2022 |
| UNI-RNA | SC, FP | Unsupervised learning | RNA | Transformer | A context-aware foundation model pre-trained on an unprecedented scale of RNA sequences unraveling evolutionary and structural information | Wang *et al.*, 2023 |
| scFoundation | FP | Unsupervised learning | scGenomics | Transformer | An extensive single-cell foundation model pre-trained on a dataset of over 50 million single-cell data points with 100 million parameters | Hao *et al.*, 2023 |
| scHyena | FP | Unsupervised learning | scGenomics | Transformer | A full-length scRNA-seq analysis in the brain by a linear adaptor layer and a bidirectional Hyena operator without losing raw data information | Oh *et al.*, 2023 |
| scBERT | FP | Unsupervised learning | scGenomics | Transformer | Use self-supervised learning on large-scale unlabeled scRNA-seq data to improve the model's generalizability and overcome the batch effect | Yang *et al.*, 2022 |
| ProtST | FP, MP | Unsupervised learning | Protein, biomedical text | CNN \| Transformer | A pre-trained framework with three tasks of both protein and biomedical text to boost protein sequence understanding | Xu *et al.*, 2023 |